\documentclass[a4paper,fleqn,usenatbib]{mnras}
\usepackage{newtxtext,newtxmath}

\usepackage[T1]{fontenc}
\usepackage{ae,aecompl}

\usepackage{graphicx}	

\usepackage{amsmath}	
\usepackage{amssymb}	

\title[Spot group lifetimes]{Constraints on sunspot group lifetimes from far-side Sun-as-a-star helioseismology with BiSON}

\author[W. J. Chaplin et~al.]{William J. Chaplin$^{1}$\thanks{E-mail: w.j.chaplin@bham.ac.uk}, Rachel Howe$^1$, Sarbani Basu$^2$, Yvonne Elsworth$^1$,
\newauthor Steven J. Hale$^1$, Emily J. Hatt$^1$, Eleanor Murray$^1$ and Martin B. Nielsen$^1$\\
$^1$School of Physics and Astronomy, University of Birmingham, Birmingham, B15 2TT, United Kingdom\\ $^{2}$Department of Astronomy, Yale University, PO Box 208101, New Haven, CT, 065208101, USA}

\date{Accepted XXX. Received YYY; in original form ZZZ}

\pubyear{2025}

\begin{document}
\label{firstpage}
\pagerange{\pageref{firstpage}--\pageref{lastpage}}
\maketitle

\begin{abstract}

Frequencies of low-degree solar p modes are sensitive to activity over the entire Sun, including the unobservable far-side hemisphere. When frequency shifts extracted from week-long BiSON datasets are fitted to a linear combination of observed near-side activity and a far-side proxy made from the near-side measures shifted by half the solar rotation period, the solution favours a slightly higher weighting from the far-side contribution. Here, we demonstrate that this unphysical mismatch is due to the inherent inaccuracy of the far-side proxy, which fails to capture active regions that evolve fully on the solar far side, or that evolve (or have evolved) significantly as they rotate off (or onto) the visible disc. By simulating the evolution of sunspot group areas over time, which act as a suitable measure of solar activity, we show that the solution is sensitive to the lifetime of the activity. Assuming an underlying mapping from maximum group areas $A_{\rm max}$ (measured in millionths of the solar hemispheric area, MSH) to group lifetimes $\tau$ (measured in days) of the form $\tau = \alpha A_{\rm max}$, we find that $\alpha \simeq 0.025^{+0.055}_{-0.016}\,\rm d\,MSH^{-1}$ gives results consistent with the BiSON finding. This is to be compared with the value of $\alpha = 0.1\,\rm d\,MSH^{-1}$ implied by the well-known Gnevyshev-Waldmeier rule. While our best-fitting $\alpha$ maps to an average group lifetime of $\tau \simeq 5^{+10}_{-3}\,\rm d$, the best-fitting distribution includes a reasonable fraction of groups with lifetimes longer than the solar rotation period, which is essential to reproducing the mismatch. 

\end{abstract}




\section{Introduction}
\label{sec:intro}

It is now well established that the frequencies of the Sun's p-mode oscillations vary systematically over time, in response to changing levels of solar activity as the Sun's 11-yr cycle waxes and wanes (e.g., see \citealt{Chaplin14, Howe18, Broomhall2022, Garcia24, Baird24}; and references therein). The mode frequencies therefore act as a probe of the sub-surface changes that are responsible for driving their variation \citet{Basu21}. In \citet{howe25}, we showed it is possible to extract robust activity-driven frequency shifts of low-degree solar p modes from time-series segments as short as one week, using Sun-as-a-star Doppler velocity data collected by the Birmingham Solar-Oscillations Network (BiSON; \citealt{Chaplin1996,Hale2016}). Taking advantage of the fact that this segment duration is significantly shorter than the solar rotation period, we demonstrated that the measured frequency shifts of these low-degree modes correlate more strongly with proxies of the Sun's global activity that include a contribution to mimic activity on the unobserved solar far-side. These are truly global modes that are sensitive to the entire Sun and hence are affected by far-side features. 

When fitted to a model comprising a linear combination of the observed near-side activity, and a proxy for the far-side activity made from the observed near-side measures suitably shifted in time, the fit favoured a mix of 59\,\% of the far-side proxy and 41\,\% of the observed near-side proxy. We speculated that this significant (and unphysical) departure from the expected 50\,\%:50\,\% mix was due to the finite lifetime of the magnetic activity to which the mode frequencies respond, i.e., activity that emerged and decayed out of view on the far-side, or evolved significantly after rotating into or out of view on the near-side, had rendered the constructed far-side proxy inaccurate. Here, we demonstrate that this is indeed the case: the best-fitting coefficient describing the mix of near- and far-side proxies is sensitive to the lifetime of the activity, and we are able to place constraints not only on the mean lifetime of the activity to which the p modes are sensitive, but also how the areas of active regions -- specifically sunspot group areas -- map to lifetimes. Our conclusions and insights come from simulations of sunspot group areas, which we adopt as measures of the Sun's activity. Group areas constitute reasonably objective, robust measures that are easier to simulate than, say, sunspot numbers or the Sun's 10.7-cm radio flux. Literature data are available on the distribution of group area sizes and their evolution over time, which can be readily captured in simulations.

The layout of the rest of the paper is as follows. In Section~\ref{sec:sim} we describe how we built a simulated time-domain record of sunspot group areas covering the surface of a rotating, synthetic Sun, and compare the resulting simulated records of observed, disc-integrated group areas with real observations. In Section~\ref{sec:rep} we then use our simulated whole-Sun-integrated group signal as a proxy of the p-mode frequency shifts, in order to replicate the analyses performed by \citet{howe25}. The results of the full analysis are presented in Section~\ref{sec:res},  where we show how we were able to establish the sensitivity of the best-fitting mix coefficient to the lifetime of the simulated activity. We finish by drawing our main conclusions in Section~\ref{sec:conc}.

\section{Simulation of sunspot group areas}
\label{sec:sim}

\begin{figure*}
\centering
\includegraphics[width=0.5\textwidth]{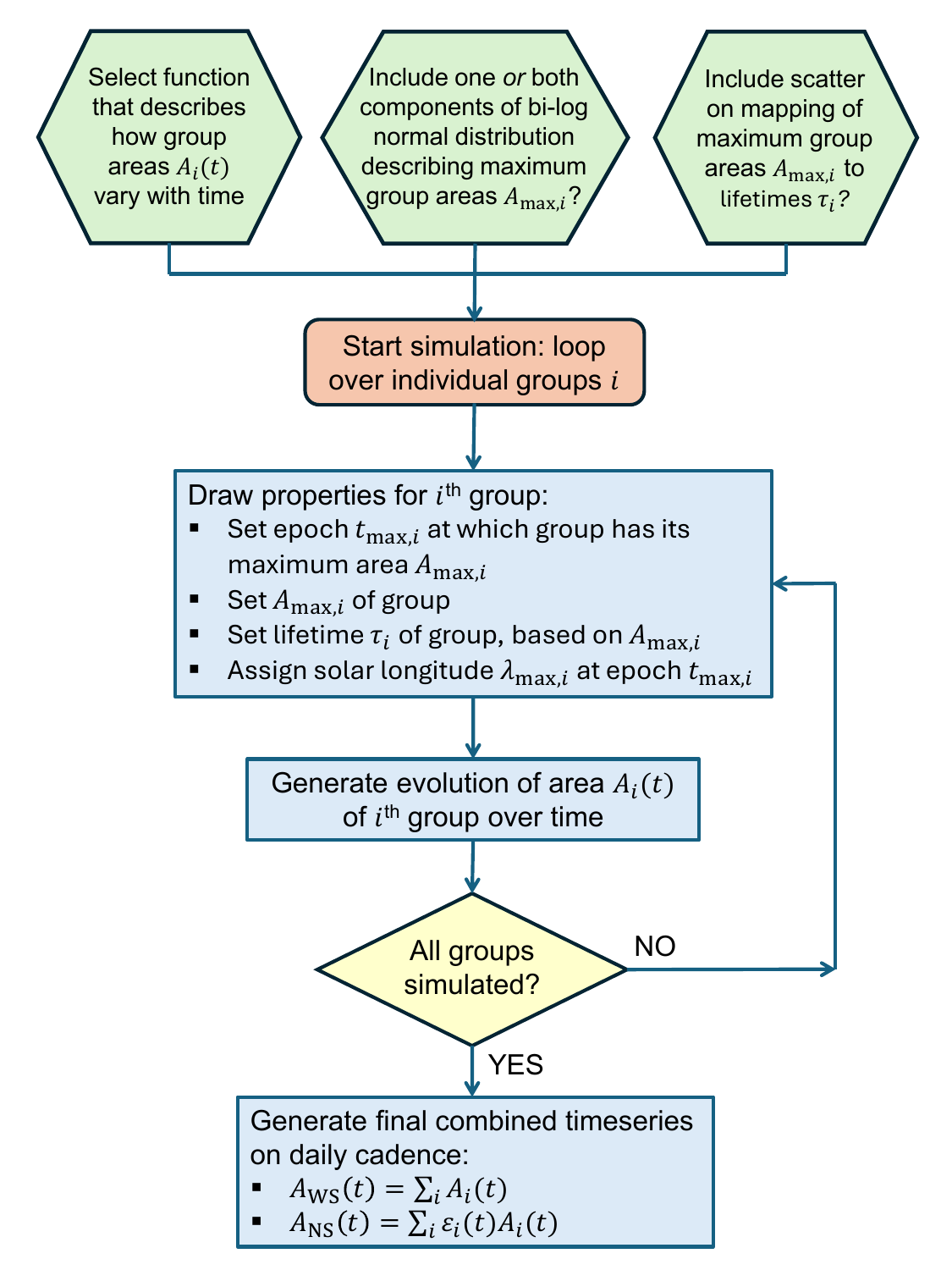}
        \caption{Schematic flow chart covering the key steps in the simulations of sunspot group areas (see text for details of each step).}
	  \label{fig:flow}
\end{figure*}

Our results are based on time-domain simulations of sunspot group areas, covering the surface of a synthetic Sun. Each realization spanned a simulated duration of $\simeq 32\,\rm yr$, to mimic the slightly less than three full 11-yr Schwabe cycles covered by the BiSON results presented in \citet{howe25}. The simulations are built from a granular level, by simulating the variation over time of the area of individual sunspot groups. A linear combination at each epoch of all individual group records then provided measures of the total sunspot area. Data were generated on a 1-d cadence to match the standard daily reporting of total spot group areas. Fig.~\ref{fig:flow} is a schematic flow chart covering the key steps in the simulations, which we now go on to discuss in detail. In what follows, individually simulated spot groups are tagged with the integer $i$.


\subsection{Underlying group properties}
\label{sec:props}

\subsubsection{Maximum group area $A_{{\rm max},i}$}
\label{sec:areadis}

There is now evidence in the literature that the distribution of maximum group areas may be described by two populations, of smaller and larger features (e.g., see \citealt{nagovitsyn16, nagovitsyn21}, and references therein). Each group was assigned its maximum area $A_{{\rm max},i}$, calibrated in millionths of the area of a solar hemisphere (MSH), by drawing areas randomly from the bi-log-normal distribution presented in \citet{nagovitsyn21}. This comprises one distribution of smaller features, having a peak area of $16\,\rm MSH$ and a FWHM of 0.47\,dex; and a second distribution of larger features, having a peak area of $219\,\rm MSH$ and a FWHM of 0.43\,dex. (These areas correspond to radii for circular features of $\simeq2,800$ and 10,300\,km, respectively.) The parameters describing the second distribution above are in reasonable agreement with the results of \citet{forgacs21}, who studied group areas and lifetimes but with an area threshold cut that excluded smaller features. This meant their results essentially reproduced the second distribution of \citet{nagovitsyn21}. In our simulations, we tested the impact on the results of including one or both of the area distributions.

\subsubsection{Group lifetime $\tau_i$}
\label{sec:life}

The relationship between group areas and their lifetimes has been well studied, dating back to the empirical Gnevyshev-Waldmeier rule (\citealt{Gnevyshev38, Waldmeier55}). This asserted a linear relation between $A_{\rm max}$ (as expressed in MSH) and lifetime $\tau$ (expressed in d) of the form $A_{\rm max} = 10\tau$. Subsequent studies have found that sunspot groups show behaviour that is broadly consistent with the Gnevyshev-Waldmeier rule, albeit with some scatter and uncertainty on the linear calibration constant of $10\,\rm MSH\,d^{-1}$ (e.g., see \citealt{petrovay97, henwood10, nagovitsyn16, forgacs21}). For example, \cite{nagovitsyn16} and \citet{forgacs21} found that constants of, respectively, 13 and $20\,\rm MSH\,d^{-1}$ were a better fit to observations (albeit again noting the presence of scatter about the linear relation). 

In our base-level simulations, we fixed group lifetimes according to
\begin{equation}
    \tau_i = \alpha A_{{\rm max},i},
    \label{eq:tau}
\end{equation}
with $\tau_i$ again in units of days, and $A_{{\rm max},i}$ again in units of MSH. Strict adherence to the Gnevyshev-Waldmeier rule would imply a value for our linear coefficient of $\alpha = 0.1\,\rm d\,MSH^{-1}$; whilst the results from \cite{nagovitsyn16} and \citet{forgacs21} correspond to values of $\alpha$ of, respectively, $0.08\,\rm d\,MSH^{-1}$ and $0.05\,\rm d\,MSH^{-1}$. We also note that \citet{Tlatov23} found $\alpha \simeq 0.02$ to $0.03\,\rm d\,MSH^{-1}$ from analysing data on individual sunspots.

The use of the simple scaling in Equation~\ref{eq:tau} allowed us to straightforwardly test the sensitivity of the results to $\alpha$. However, cognisant of the fact that real observations show scatter in this relationship -- though how much of this is intrinsic and how much is based on measurement uncertainty is unclear -- we also ran simulations where we included intrinsic scatter about Equation~\ref{eq:tau}. Our implementation was guided by the results in \citet{forgacs21}. In a plot of measured group lifetimes versus measured maximum group areas, they found that the results were bounded approximately by two lines. The Gnevyshev-Waldmeier rule $\tau = 0.1A_{\rm max}$ constituted a rough upper bound, while the relation $\tau = 1.25A_{\rm max}^{1/3}$ fixed a lower bound. At any given $A_{{\rm max},i}$, we used the distance in $\tau$ between these two lines to define a 6$\sigma$ range. We then added random deviates drawn from a Gaussian distribution having a standard deviation of $\sigma$ to the lifetime fixed by Equation~\ref{eq:tau}, to give scattered values for $\tau_i$.

\subsubsection{Epoch $t_{{\rm max},i}$ of maximum group area}

The epoch $t_{{\rm max},i}$ at which a spot group showed its maximum area $A_{{\rm max},i}$ was drawn from a random distribution. The probability of a group having its peak size at a given epoch rose or fell with time across the simulated period to match the simulated rising or falling phases of each simulated cycle. To achieve this, we used the inverse transform sampling method, drawing uniformly distributed random samples from the inverse of the cumulative distribution function describing the simulated solar cycle variation in time. This probability distribution was calibrated so that the combined spot group area of the simulated solar near-side signal was a reasonable match to real observations, as captured by the consolidated catalogue of \citet{mandal20}. This typically required $\approx 1000$ to 1500 simulated spot groups on each rising or falling cycle phase. We tested the impact on the results of using either a sinusoidal or a simplified triangular function to represent the underlying variations of spot group numbers over the simulated cycles. This choice did not significantly affect the results or conclusions. Neither did changing the amplitudes of different simulated cycles relative to one another in the simulated 32-yr period. However, we note that we did not include additional complexity, e.g., changes to the proportion of smaller and larger group areas as a function of activity in one cycle or from one simulated cycle to the next (e.g., see \citealt{Lefevre2011}). 
In what follows, we present results from using simple triangular functions for the cyclic variation.

\subsubsection{Functional dependence of group area $A_i(t)$ over time $t$}

The time-dependent area $A_i(t)$ of each group was assumed to follow one of three functions in time. Here, we again took a lead from the analysis of observed sunspot group areas presented by \citet{forgacs21}. They found that an asymmetric Gaussian function could be used to fit the observed variations of different group areas in time. In some of our simulations we therefore adopted their asymmetric function (Equation~1 in their paper), fixing the associated asymmetry parameter for each simulated group by randomly drawing samples from a normal distribution that matched approximately their distribution of fitted asymmetries (see their Fig.~9). We also ran simulations assuming two other functional forms: a simpler Gaussian function, and a double-exponential function. For each case, the FWHM of the adopted distribution corresponded to the lifetime, $\tau_i$.

\begin{figure}
\centering
\includegraphics[width=0.5\textwidth]{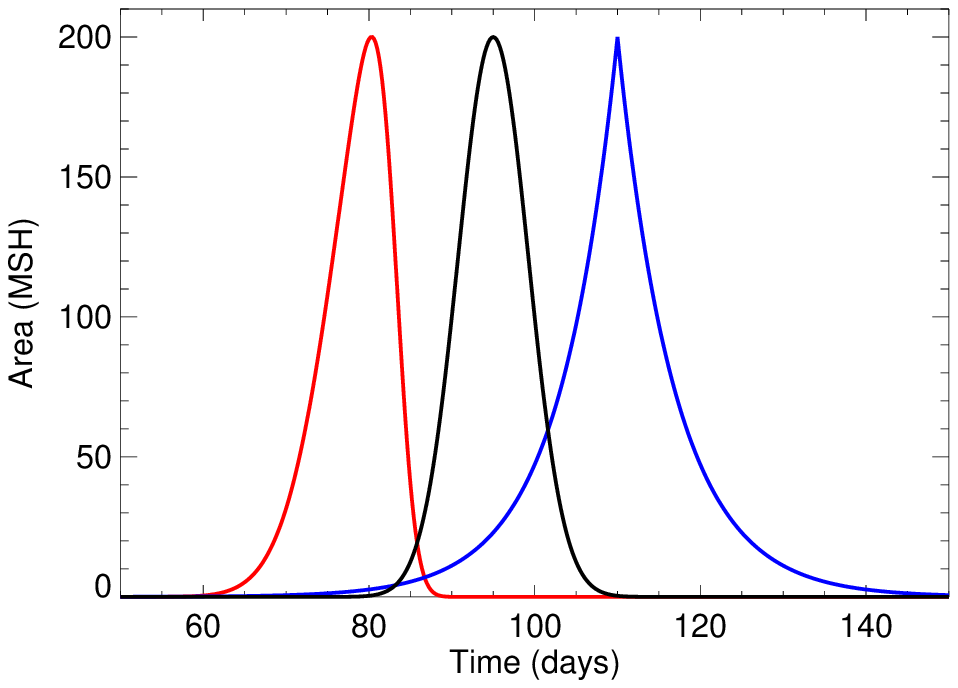}
        \caption{$A_i(t)$ for a simulated group of $A_{{\rm max},i}=200\,\rm MSH$ and $\tau_i=10\,\rm d$. Variation shown for the asymmetric function in red, Gaussian function in black, and double-exponential function in blue. The $t_{{\rm max},i}$ are offset for clarity.}
	  \label{fig:look1}
\end{figure}

To illustrate the different functional forms, Fig.~\ref{fig:look1} shows $A_i(t)$ for a simulated group of $A_{{\rm max},i}=200\,\rm MSH$ and $\tau_i=10\,\rm d$. The variation for a moderately strong asymmetric function is plotted in red\footnote{This has an asymmetry parameter of $n=-3.5$ following the definition in \citet{forgacs21}, and is similar to the real data examples shown in their Fig.~3.}, for a Gaussian function in black, and for a double-exponential function in blue. Note $t_{{\rm max},i}$ is offset for clarity.


\subsubsection{Solar longitude $\lambda_{{\rm max},i}$ of group at epoch $t_{{\rm max},i}$}

Each group was deposited on the simulated solar surface with the epoch $t_{{\rm max},i}$ of its maximum area associated to a random solar longitude $\lambda_{{\rm max},i}$, drawn from a uniform distribution with the zero point lying on the central meridian (i.e., the Stonyhurst definition of longitude). As such, our simulations did not include preferred bands of active longitudes. We assumed all spots rotated on the simulated solar surface with a period of $P_{\rm rot} = 26\,\rm \,d$. Including the latitudinal variation of the spot groups over time, and the associated dependence of the rotation period, did not significantly impact the results.

\subsection{Construction of simulated timeseries}

With $t_{{\rm max},i}$, $A_{{\rm max},i}$, $\tau_i$ and $\lambda_{{\rm max},i}$ fixed for all spot groups, we then calculated records of how the area and solar longitude of each group varied with time. It was then a simple matter to produce composite timeseries, representing the combined spot group areas.

First, a simple linear combination at each epoch of all individual group records provided a measure over time of the total spot group area across the entire solar surface. In what follows, we refer to this as the total whole-Sun (WS) area, i.e., 
 \begin{equation}
 A_{\rm WS}(t) = \sum_i A_i(t).
 \label{eq:WS}
 \end{equation}
The simulated near-side signal was then constructed by combining the contributions of individual spot groups at observable longitudes only. We refer to this as the total near-side (NS) area. It is given by:
\begin{equation}
 A_{\rm NS}(t) = \sum_i \varepsilon_i(t) A_i(t),
 \label{eq:NS}
 \end{equation}
where $\varepsilon_i(t)$ captures the visibility of each group over time according to:
\begin{equation} 
 \mbox{$\varepsilon_i(t)=$} \left\{
 \begin{array}{lll} 
 \mbox{$\sin \lambda_i(t)$}&
 \mbox{Near-side group, $3\pi/2 \le \lambda_i(t) < \pi/2$;}\\
 \mbox{0}&
 \mbox{Far-side group, $\pi/2 \le \lambda_i(t) < 3\pi/2$} 
 \end{array} \right.
 \label{eq:winT} 
 \end{equation} 
Here, $\lambda_i(t)$ is the longitude of each group at time $t$, calculated from
 \begin{equation}
 \lambda_i(t) = \left[ \lambda_{{\rm max},i} + 2\pi/P_{\rm rot} \left( t - t_{{\rm max},i} \right) \right]\,{\rm mod}\, 2\pi,
 \end{equation}
and the factor $\sin \lambda_i(t)$ in Equation~\ref{eq:winT} accounts for the area projection as groups move across the visible disc.


\begin{figure*}
\centering
\centerline{\includegraphics[width=0.5\textwidth]{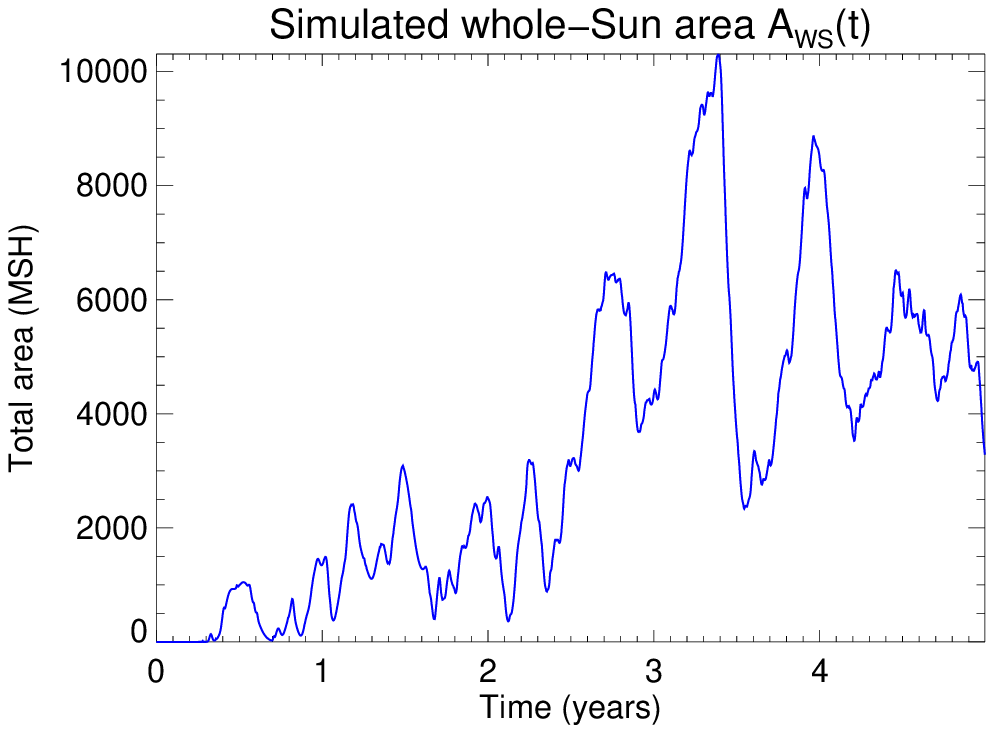}}
\centerline{\includegraphics[width=0.5\textwidth]{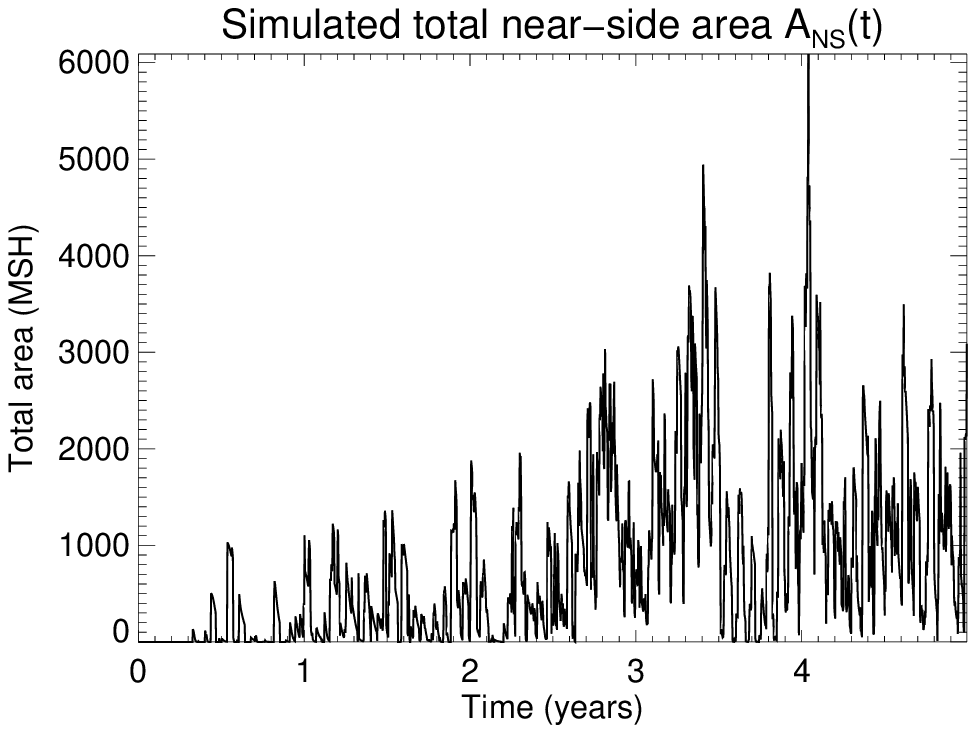}
            \includegraphics[width=0.5\textwidth]{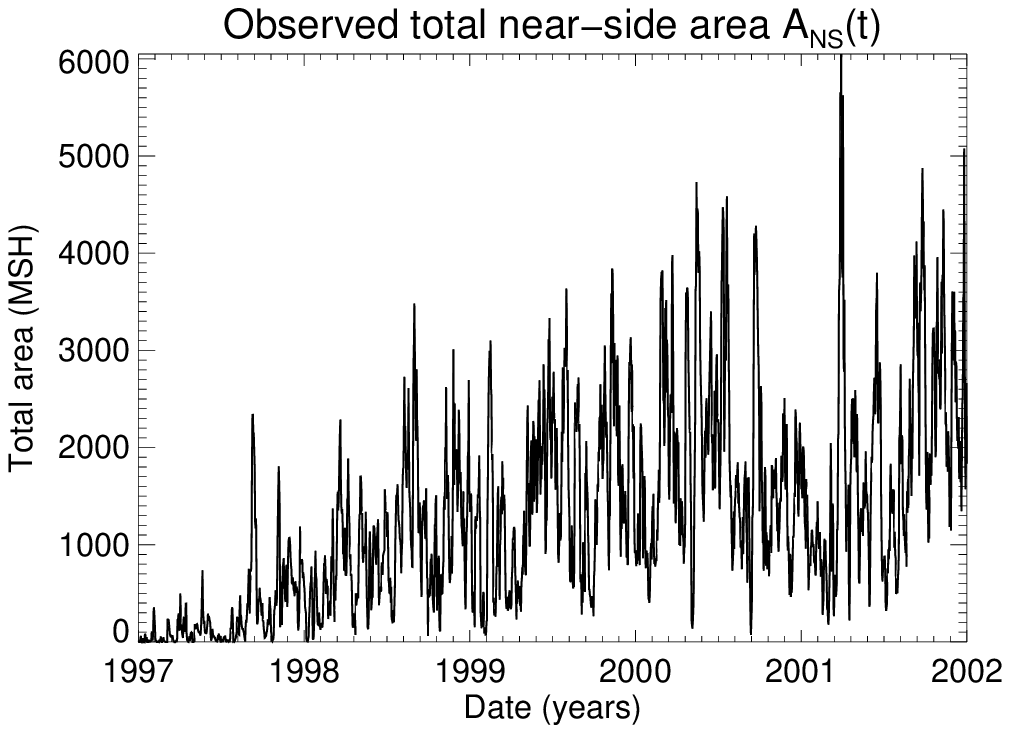}}
        \caption{Top panel: Simulated daily record of the whole-Sun group area from a single 5-yr realization of our simulations, in units of millionths of the area of a solar hemisphere (MSH). Bottom left-hand panel: corresponding simulated total near-side area, as made from the whole-Sun record in the top panel. Bottom right-hand panel: Real, observed (i.e., near-side) total areas for comparison, showing a 5-yr record from the rising phase of Cycle~23.}
	\label{fig:life1}
\end{figure*}


Fig.~\ref{fig:life1} shows simulated daily records from a 5-yr segment, spanning the rising phase of a simulated cycle from a single realization of our simulations. The top panel plots the simulated whole-Sun area, while the bottom left-hand panel plots the simulated total near-side area. This realization contained 1000 groups over the plotted segment; lifetimes derived from the simple relation $\tau =0.05A_{\rm max}$; and both components of the bi-log-normal distribution used to describe the underlying distribution of group areas. For comparison, the bottom right-hand panel shows real observed daily records over a 5-yr period on the rising phase of cycle~23, from the \citet{mandal20} catalogue of areas. The general appearance and signal characteristics of the simulated near-side areas show a striking resemblance to real data. The simulated whole-Sun record is much smoother, indicating that (not surprisingly) the ragged nature of the near-side signal is a result of rotational modulation of the area signal.

\section{Replicating the BiSON analysis}
\label{sec:rep}

With the simulations established, we then sought to explain the result in \citealt{howe25}, i.e., why does a fit of the BiSON frequencies to a model comprising the observed near-side proxy, plus a proxy for the far-side activity, favour a slightly higher weighting from the far-side contribution? 

\begin{figure}
\centering
\includegraphics[width=0.5\textwidth]{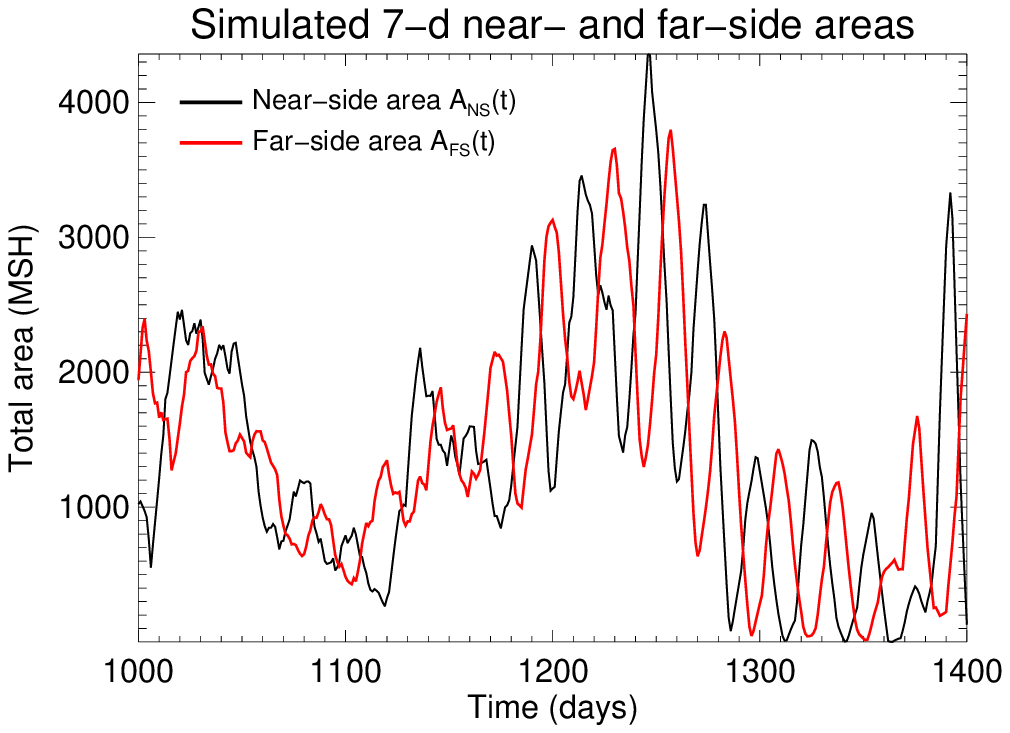}
        \caption{Zoom in time of the simulated total near-side area from Fig.~\ref{fig:life1} (black), averaged over 7\,d, and the resulting far-side proxy (red) constructed from it.}
	  \label{fig:far1}
\end{figure}

We followed \citet{howe25} and analysed the data in 7-d segments. First, daily total near-side and total whole-Sun data produced by each simulation were averaged on a 7-d timescale. Next, far-side activity proxies were constructed from 7-d averages of the near-side activity over epochs $P_{\rm rot}/2 \equiv 13\,\rm d$ earlier and $13\,\rm d$ later than the observed 7-d period. Fig.~\ref{fig:far1} shows a zoom in time of the simulated total near-side area from Fig.~\ref{fig:life1} (black line) and the resulting far-side proxy (red line) constructed from it. As was noted by \citet{howe25}, these two signals are often in anti-phase. 


Next, assuming the total whole-Sun area provides a good proxy for the total magnetic activity to which the modes are most sensitive, we then constructed a proxy of the p-mode frequency shift data from the 7-d averaged, simulated total whole-Sun area. First, we scaled the 7-d whole-Sun timeseries $\left< A_{\rm WS} (t) \right>$ by a factor $k$ to have a range comparable to the observed p-mode frequency shifts (e.g., see \citealt{Howe18, chaplin19}). We then added a random Gaussian deviate $X(t)$ to each element of the resulting scaled timeseries, to simulate the typical uncertainty observed on the real 7-d BiSON frequency shifts, having a standard deviation of $\sigma = 0.09\,\rm \mu Hz$. Our proxy of the frequency shift data was therefore:
 \begin{equation}
  \left< \delta\nu (t) \right> = k \left< A_{\rm WS} (t) \right> + X(t),
 \end{equation}
with $X(t) \sim \mathcal{N}(0,\sigma)$. For any given realization of our simulations, we then followed \citet{howe25} and fitted a linear model of the form
\begin{equation}
  \left< \delta\nu (t) \right> = S [( 1-\beta) \left< A_{\rm NS} (t) \right> + \beta \left< A_{\rm FS} (t) \right>] + c,
  \label{eq:fit}
\end{equation}
where $\left< A_{\rm NS} (t)\right>$ and $\left< A_{\rm FS} (t)\right> $ are the 7-d averaged near- and far-side areas respectively, $\beta$ is the mix parameter controlling the relative contributions of these areas, $S$ is a sensitivity term, and $c$ is a constant.

For a given set of input parameters -- specifying the distribution of group areas and lifetimes -- we repeated the fitting analysis on independent realizations of the simulations, thereby seeking to establish the range of parameters that would replicate the \citet{howe25} results. 

\section{Results}
\label{sec:res}

\subsection{Repeating the BiSON analysis with sunspot group areas}
\label{sec:resbison}

Our analyses on the real BiSON data in \citet{howe25} used the 10.7-cm radio flux as the activity proxy. It was from these fits that we obtained $\beta = 0.59 \pm 0.04$, i.e., a best-fitting model comprising 59\,\% of the far-side proxy and 41\,\% of the observed near-side proxy. A similar mix was returned when using the sunspot number record as the activity proxy. 

Since our simulations here are based instead on sunspot group areas, we repeated the BiSON fits using the real group area data as the activity proxy to check we obtained similar results. We constructed the required 7-d near- and far-side averages using the daily total areas compiled by \citet{mandal20}. A fit of the 7-d BiSON frequency shifts to these proxy averages yielded a best-fitting estimate of the coefficient $\beta$ of $0.60 \pm 0.04$, in good agreement with the results returned using the other proxies. This serves as our baseline comparison for results obtained from the simulated data, which we now go on to discuss.

\subsection{Results from the simulations}
\label{sec:ressim}

\begin{figure*}
\centering
\includegraphics[width=0.65\textwidth]{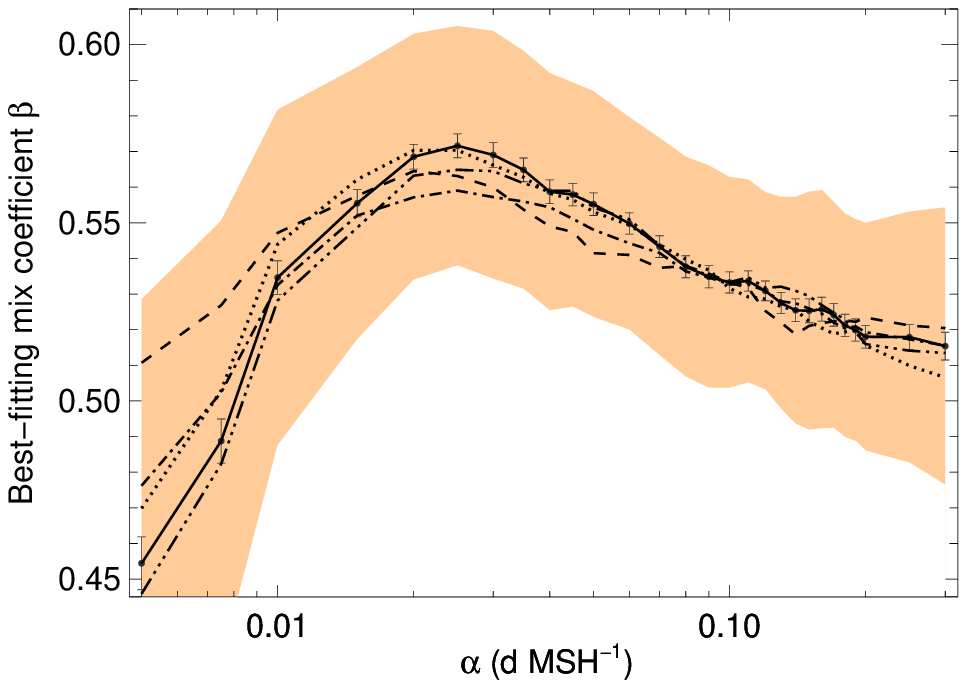}
\includegraphics[width=0.65\textwidth]{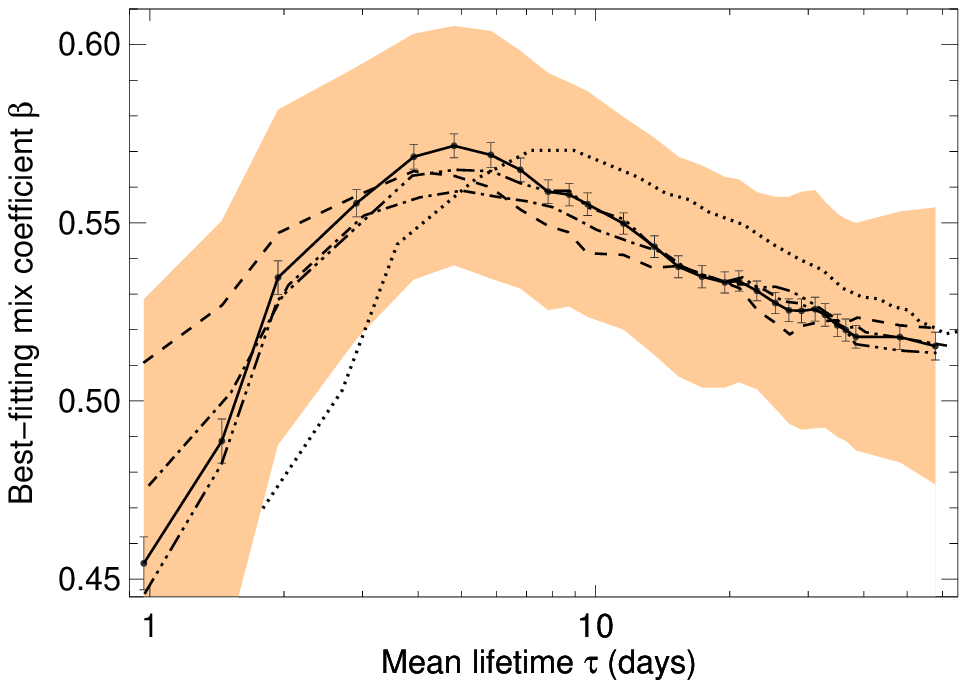}
\caption{Top panel: mean best-fitting coefficient $\beta$ given by fits of different sets of simulated data to Equation~\ref{eq:fit}, plotted as a function of $\alpha$. Bottom panel: same, but using as the independent variable the mean lifetime $\tau$ that results from each selected value of $\alpha$. Results from the base simulation set S1 (see main text for details) are plotted with the solid line. The error bars show the 1-sigma error on the mean of 100 realizations of the base simulations, whilst the coloured, shaded region bounds the region spanning the standard deviation over each set of 100 realizations. The standard deviations are (by design) similar to the uncertainty on the fit to the real BiSON dataset. Results for the other sets of simulations are rendered using different line styles: S2 as a dotted line, S3 dashed, S4 dot-dashed, and S5 dot-dot-dot-dashed (see text for details). Each set of results is again an average over 100 independent noise realizations. They have very similar errors and standard deviations to the set S1, and hence for visual clarity we plot errors for one set only.}
\label{fig:resall}
\end{figure*}

The top panel of Fig.~\ref{fig:resall} shows the mean best-fitting coefficient $\beta$ given by fits of different sets of simulated data to Equation~\ref{eq:fit}, as a function of $\alpha$. The bottom panel instead uses as the independent variable, the mean lifetime $\tau$, that results from each selected value of $\alpha$. Results for different set of simulations are rendered using different line styles, with each set of results being an average over 100 independent noise realizations. 

The base results rendered using the solid line -- which we refer to as set S1 -- are from simulations that included both components of the bi-log normal area distribution; used the simple linear mapping $\tau = \alpha A_{\rm max}$ from maximum area $A_{\rm max}$ to lifetime $\tau$ described by Equation~\ref{eq:tau}; and modelled the variation of group areas over time using a Gaussian function. The error bars show the 1-sigma uncertainty on the mean of 100 realizations of the base simulations. The coloured shaded region spans the standard deviation over each set of 100 realizations; this spread is (by design) similar to the uncertainty on $\beta$ shown by the fit to the real BiSON data. The errors and spread given by fits to the other sets of simulations are of a similar size to those shown for the base simulations. 

Moving on to the other sets, results rendered with the dotted line (S2) were based on the same inputs as the first set above, apart from the underlying area distribution where only the second larger-area component of the bi-log normal distribution was included, and the smaller-area component was excluded. The next sets of simulations adopted different functional forms to describe the change of group areas over time. Results shown with the dashed line (S3) were based on inputs as per the first, base set, but with the two-sided exponential function substituted for the simple Gaussian function; whilst results shown with the dot-dashed line (S4) were instead based on use of the asymmetric Gaussian function with a distribution of asymmetry parameters applied across the simulated groups. Finally, results shown using the dot-dot-dot-dashed line (S5) are from simulations as per set S1, but with normally distributed scatter added to the underlying $\tau = \alpha A_{\rm max}$ relation.


We find that $\alpha \simeq 0.02$ to $0.03\,\rm d\, MSH^{-1}$ returns a best-fitting mix parameter $\beta$ similar to that given by fits to the real BiSON data (formally within $1\sigma$). All sets show the same trend, having a clear maximum at these values of $\alpha$. Higher and lower values of $\alpha$ -- which map to longer and shorter lifetimes, respectively, as shown in the right-hand panel -- return smaller values of $\beta$. This result is not surprising and may be understood as follows. Very long lifetimes imply a signal that tends to being stationary over time, which notionally favours a 50\,\%:50\,\% mix of near- and far-side signal; while at the other extreme, lifetimes significantly shorter than the solar rotation period will also imply a signal that tends to having (on average) stationary properties, unaffected by the impact of rotational modulation\footnote{Strictly speaking, and with no informed priors, all values of $\beta$ are equally likely for both these extreme cases. But this would not necessarily produce a physically meaningful result consistent with what we actually see on the Sun, i.e., a very high or very low value of $\beta$ would imply activity confined almost exclusively to one hemisphere, and would certainly not reflect a realistic distribution in longitude of spot groups across the activity cycle.}. It is also worth noting that results for small $\alpha$ will be affected by the finite 1-d resolution of the data. 

\begin{figure*}
\centering
\centerline{
\includegraphics[width=0.5\textwidth]{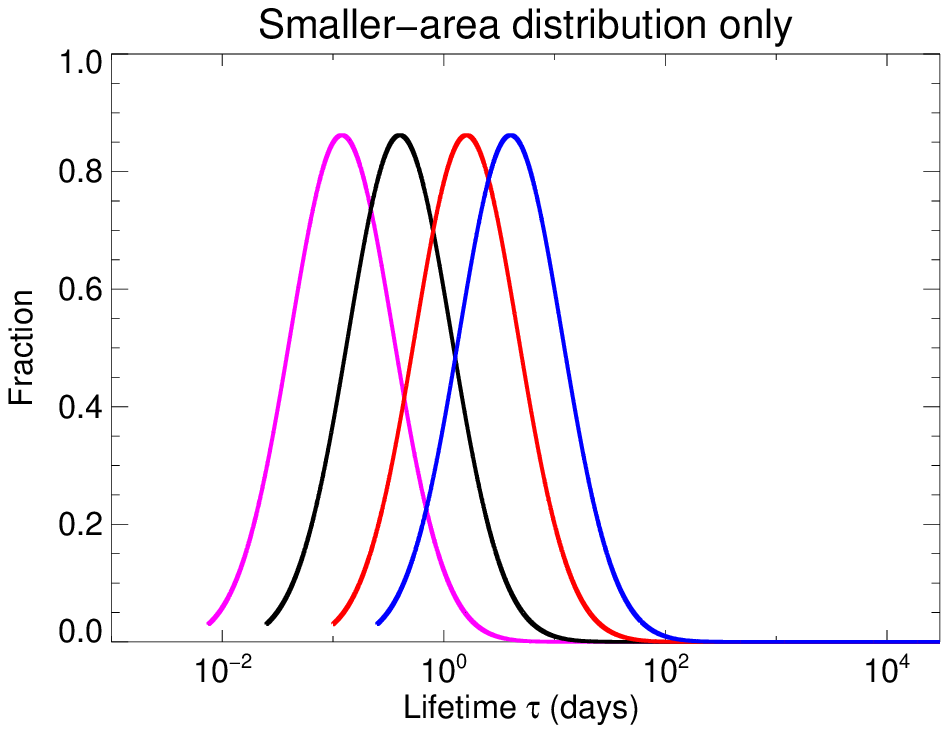}
\includegraphics[width=0.5\textwidth]{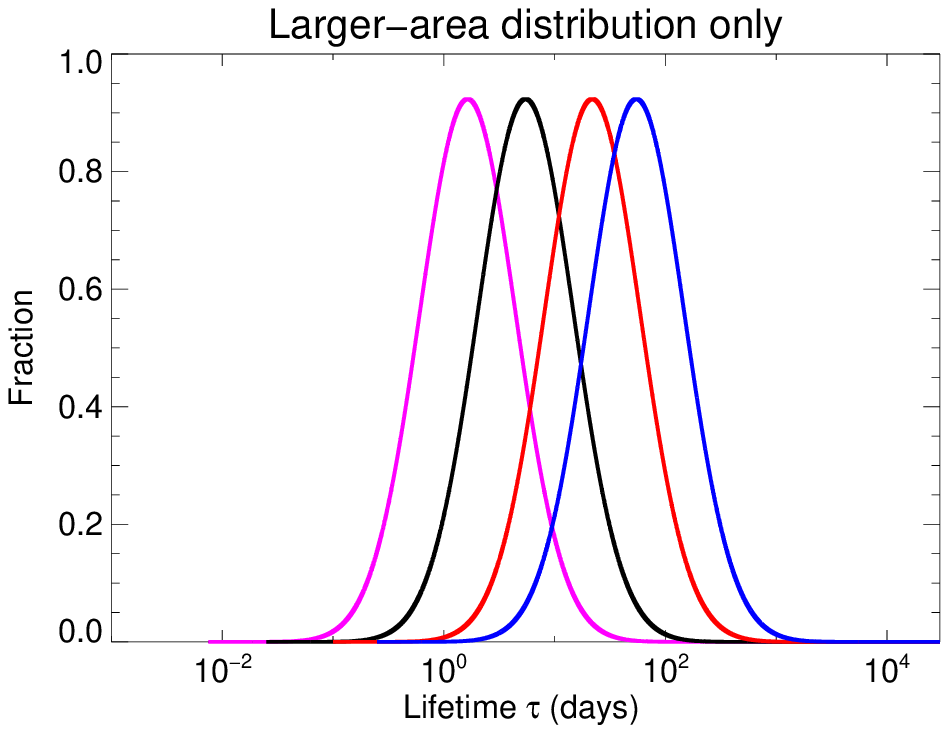}}
\centerline{
\includegraphics[width=0.5\textwidth]{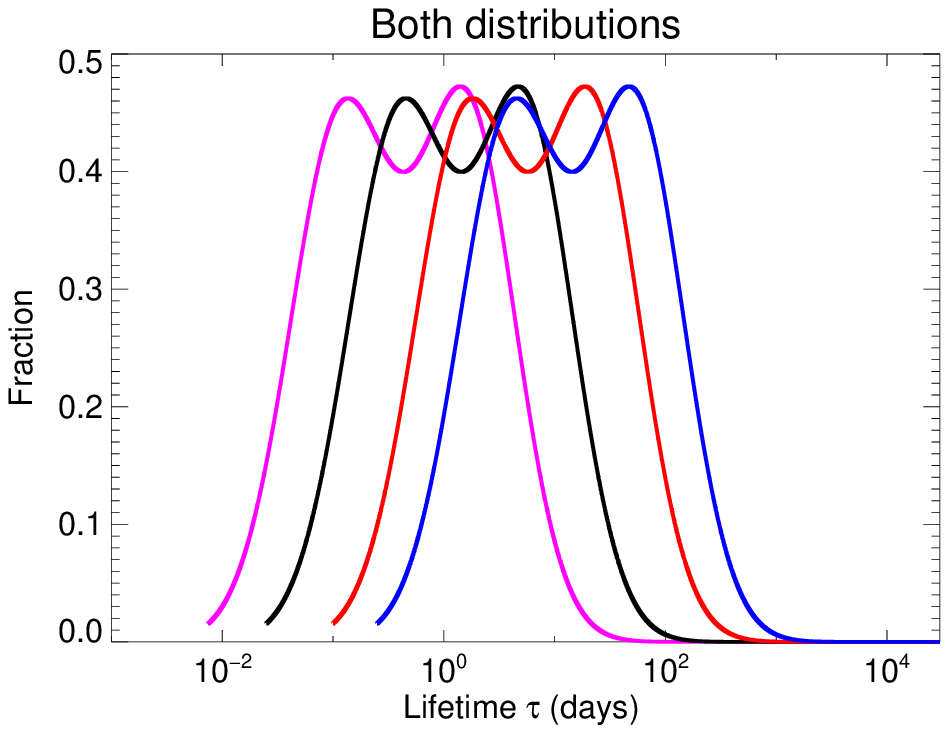}}
\caption{Underlying distributions of group lifetimes $\tau$ given by Equation~\ref{eq:tau} for values of $\alpha = 0.0075$ (pink), 0.025 (black), 0.100 (red) and $0.200\,\rm d\,MSH^{-1}$ (blue), for: the smaller-area distribution only (top left-hand panel); the larger-area distribution only (top right-hand panel), and when both distributions are included (bottom panel).}
\label{fig:distlife}
\end{figure*}

So, it is when the underlying distribution contains a mix of lifetimes both longer and shorter than the solar rotation period -- with, crucially, there being some fraction of the longer lifetime groups -- that we replicate the real result using the BiSON data. To reinforce this point, the bottom panel of Fig.~\ref{fig:resall} highlights the underlying difference between the simulations that included both components of the bi-log normal area distribution (i.e., sets S1, S3, S4 and S5), and the one case that did not (set S1, rendered with the dotted line). Here, the results are now plotted with the average lifetime as the independent variable. The simulation set S1 shows a longer mean $\tau$ at the same value of $\alpha$ due to the absence of the smaller-area distribution. That \emph{all} simulation sets, including set S1 that used the larger-area distribution only, show a very similar trend with $\alpha$ suggests that the smaller-area component has little influence on the results. To help understand why, Fig.~\ref{fig:distlife} shows how the underlying area distributions map to lifetimes $\tau$ through Equation~\ref{eq:tau} for values of $\alpha = 0.0075$, 0.025, 0.100 and $0.200\,\rm d\,MSH^{-1}$. We have plotted the underlying distribution of lifetimes for the smaller-area distribution only (top left-hand panel), the larger-area distribution only (top right-hand panel), and when both distributions are included (bottom panel). It is clear that at lower values of $\alpha$, the smaller-area distribution gives rise to group lifetimes that are almost all shorter than the solar rotation period, with many significantly shorter than one day. There is therefore little if any memory in the system on the timescales required for the effects of rotation to influence the results.


\begin{figure*}
\centering
\includegraphics[width=0.65\textwidth]{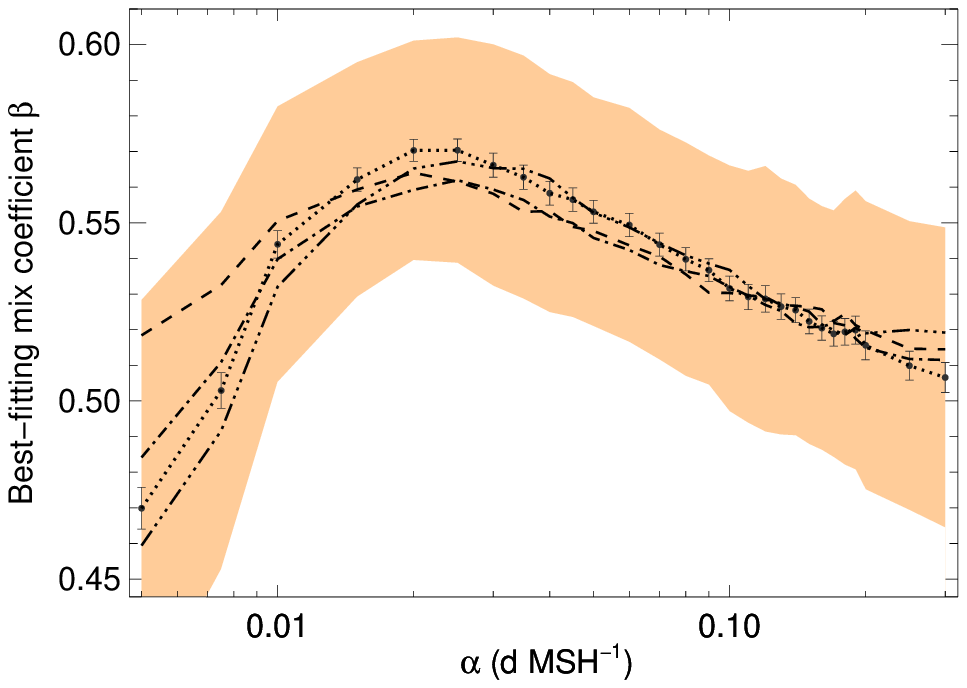}
\includegraphics[width=0.65\textwidth]{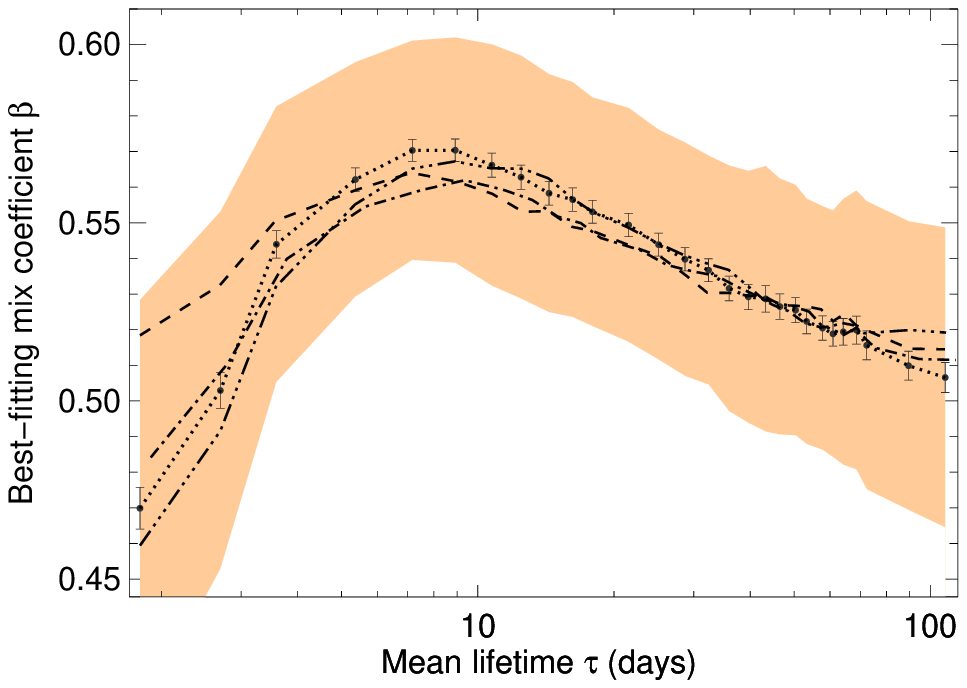}
\caption{As per Fig.~\ref{fig:resall}, but for results where \emph{all} sets of simulations included \emph{only} the larger-area component, and the smaller-area component was excluded. The mapping of line styles to different underlying simulation inputs is the same as that used in Fig.~\ref{fig:resall}. The error bars and spread are now associated with the simulation results from set S2, plotted using the dotted line.}
\label{fig:resallu}
\end{figure*}

Given the implied importance of the larger-area, longer-lifetime area distribution, in Fig.\ref{fig:resallu} we therefore show results where simulation sets S2 through S5 now included \emph{only} this component, so that the smaller-area component was excluded. The mapping of line styles to different underlying simulation inputs is the same as that used in Fig.~\ref{fig:resall}. The error bars and spread are now associated with the simulation results from the modified set S2, plotted using the dotted line. This set's results are unchanged from before.

Taking into account the standard deviation (spread) of the fits to the individual realizations (the coloured shaded regions), the maximum in $\beta$ in Fig.~\ref{fig:resallu} maps to a central value and implied uncertainties on $\alpha$ of $\simeq 0.025^{+0.055}_{-0.016}\,\rm d\,MSH^{-1}$, and on $\tau$ of $\simeq 5^{+10}_{-3}\,\rm d$. While this $\alpha$ is lower than that reported by, for example, \citet{nagovitsyn16} [a value of $0.08\,\rm d\,MSH^{-1}$] and \citet{forgacs21} [a value of $0.05\,\rm d\,MSH^{-1}$], it is nevertheless consistent within errors. It is similar to the results on individual spots obtained by \citet{Tlatov23}. The implied mean lifetime $\tau$ is also consistent, within errors, with the mode and mean of the measured distribution of group lifetimes reported by \citet{forgacs21}. It should be borne in mind that our results will reflect the population of activity to which the p-mode frequencies are most sensitive.


\section{Conclusions}
\label{sec:conc}

We have demonstrated that results obtained by \citealt{howe25} -- on fits of p-mode frequency shifts derived from week-long segments of Sun-as-a-star data to proxies of the Sun's global activity, which crucially include a contribution to mimic activity on the unobserved solar far-side -- carry information on the lifetime of the activity. That far-side contribution was constructed from the observed near-side activity, suitably shifted in time. Fits of the frequency shifts to a combination of observed near-side and inferred far-side activity strongly favour a mix of both components, but one that departs from the expected 50\,\%:50\,\% mix. Here we show that this feature of the results is a consequence of the inferred far-side proxy being an inaccurate representation of the true state of activity on the unobserved solar hemisphere, i.e., some activity will have evolved and decayed without ever appearing on the near-side, whilst other activity will evolve as it passes onto or off the visible near-side hemisphere.

Our conclusions are informed by simulations of the evolution of sunspot group areas over the waxing and waning phases of the solar cycle. They allow us to place constraints on the lifetime of these large-scale active features, and how the lifetimes of spot groups depend on their sizes. We used a realistic description of the underlying distribution of sunspot group areas, based on real observations, which may be modelled as a bi-log normal distribution containing smaller and larger-sized groups. 

When we adopt an underlying mapping of maximum group areas $A_{\rm max}$ to group lifetimes $\tau$ of the form $\tau = \alpha A_{\rm max}$, we find that $\alpha \simeq 0.025^{+0.055}_{-0.016}\,\rm d\,MSH^{-1}$ gives results consistent with those obtained by \citet{howe25} using the BiSON data. Moreover, the results are shown to have little sensitivity to the smaller-area component of the bi-log normal distribution of sunspot group areas. This component peaks at an area of $\simeq 16\,\rm MSH$ -- comparable in size to meso-granulation cells (e.g., see \citealt{Nagovitsyn25}) -- and has lifetimes for our best-fitting $\alpha$ that are almost all shorter than the solar rotation period. Our results depend almost entirely on the larger-area component. This component peaks at an area of $\simeq 219\,\rm MSH$ -- comparable in size to super-granulation cells (e.g., see \citealt{Nagovitsyn25}) -- and for our best-fitting $\alpha$ contains a reasonable fraction of groups having lifetimes longer than the solar rotation period. That fraction is crucial to matching the real BiSON results, since they depend on there being a mismatch between the inferred and actual far-side activity, which in turn demands some fraction of group lifetimes that are comparable to or longer than the solar rotation period.

\section*{Acknowledgements}

We would like to thank all those who are, or have been, associated with the Birmingham Solar-Oscillations Network (BiSON), in particular P. Pallé and T. Roca-Cortes in Tenerife and E. Rhodes Jr. and colleagues at Mt. Wilson. W.J.C., R.H., Y.E. S.J.H. and E.M. acknowledge the support of the United Kingdom Science and Technology Facilities Council (STFC) through grant ST/V000500/1. S.B. acknowledges NASA grant 80NSSC25K7669. This research has made use of NASA's Astrophysics Data System Bibliographic Services. We thank the anonymous referee for their helpful comments.

\section*{Data Availability}

 The BiSON time series analysed here is available at \url{http://bison.ph.bham.ac.uk/opendata}.


\bibliographystyle{mnras}
\bibliography{life1}


\bsp	
\label{lastpage}
\end{document}